# Scalable, reproducible, and cost-effective processing of large-scale medical imaging datasets

Michael E. Kim[a]*, Karthik Ramadass[a,b], Chenyu Gao[b], Praitayini Kanakaraj[a], Nancy R. Newlin[a], Gaurav Rudravaram[b], Kurt G. Schilling[c], Blake E. Dewey[d], Derek Archer[e,f], Timothy J. Hohman[e,f], Zhiyuan Li[b], Shunxing Bao[b], Bennett A. Landman[a,b,c,g,h], Nazirah Mohd Khairi[b]

[a]Vanderbilt University, Department of Computer Science, Nashville, TN, USA; [b]Vanderbilt University, Department of Electrical and Computer Engineering, Nashville, TN, USA; [c]Vanderbilt University Medical Center, Department of Radiology and Radiological Sciences, Nashville, TN, USA; [d]Department of Neurology, Johns Hopkins University School of Medicine, Baltimore, Maryland, USA; [e]Vanderbilt University Medical Center, Vanderbilt Memory and Alzheimer's Center, Nashville, TN, USA; [f]Vanderbilt University Medical Center, Vanderbilt Genetics Institute, Nashville, TN, USA; [g]Vanderbilt University, Department of Biomedical Engineering, Nashville, TN, USA; [h]Vanderbilt University Institute of Imaging Science, Nashville, TN, USA

## ABSTRACT

Curating, processing, and combining large-scale medical imaging datasets from national studies is a non-trivial task due to the intense computation and data throughput required, variability of acquired data, and associated financial overhead. Existing platforms or tools for large-scale data curation, processing, and storage have difficulty achieving a viable cost-to-scale ratio of computation speed for research purposes, either being too slow or too expensive. Additionally, management and consistency of processing large data in a team-driven manner is a non-trivial task. We design a BIDS-compliant method for an efficient and robust data processing pipeline of large-scale diffusion-weighted and T1-weighted MRI data compatible with low-cost, high-efficiency computing systems. Our method accomplishes automated querying of data available for processing and process running in a consistent and reproducible manner that has long-term stability, while using heterogenous low-cost computational resources and storage systems for efficient processing and data transfer. We demonstrate how our organizational structure permits efficiency in a semi-automated data processing pipeline and show how our method is comparable in processing time to cloud-based computation while being almost 20 times more cost-effective. Our design allows for fast data throughput speeds and low latency to reduce the time for data transfer between storage servers and computation servers, achieving an average of 0.60 Gb/s compared to 0.33 Gb/s for using cloud-based processing methods. The design of our workflow engine permits quick process running while maintaining flexibility to adapt to newly acquired data.

*Corresponding Author: michael.kim@vanderbilt.edu

## 1. INTRODUCTION

As technology becomes increasingly integrated into our healthcare system, we will continue to see an upward trend in the amount of medical imaging data available for scientific research.[1–4] It is imperative that these data are properly curated and processed, as larger-scale studies have shown that an increase in sample size helps us to reach more generalizable conclusions about human health.[5–8] Further, having access to large, diverse datasets helps to prevent machine learning (ML) and deep learning (DL) algorithm bias towards certain sites or populations.[9–13] Thus, it is increasingly important to be able to manage and process large-scale datasets effectively. In particular, the field of neuroimaging has surged, with several large studies available and growing.[14–23]

However, there are a few considerations for curating and maintaining such large-scale datasets. First, the technological capacity to store all data and process them in an efficient manner is essential: having data processing be too large of a bottleneck can hinder the pace of scientific research. Using only local workstations to run processing is not scalable to large, national-scale datasets due to the overall low-throughput, or, alternatively, the large overhead in buying hardware that would be required to reach reasonable processing speeds.[24,25] For distributed computing to better parallelize computation, high performance computing (HPC) clusters are options that have been used in the field of neuroimage analysis for reproducible batch processing of large datasets.[21,24,26,27] Unfortunately, HPCs are not available to every researcher and can be limited in the amount of control individual users have in setting up the processing environments. A different option available to a researcher at any institution that offers a much higher bound on potential and control is to use cloud computing resources.[24] Yet, there are still several potential concerns with using cloud resources, such as privacy, security, and especially cost, among others.[24,28–31]

Another concern is consistency of computation for reproducible science.[32,33] Relying on local installation of software or codebases can lead to differences in versions or compatibility issues that can either make previous research methods impossible to replicate or yield different results. A common method of subverting architecture and environment setup is containerization or virtualization, which also helps ensure reproducibility of code.[34–36] In the last decade, containerization has become a more popular choice due to the more lightweight nature of containers compared to virtual machines.[37] Thus, many platforms for containerization have become available, such as Linux Containers, Singularity/Apptainer, or Docker, which all provide similar functionality with differences in use cases, such as Singularity not requiring administrative OS permissions.[34] For distributed computing, individual startup of each container can be a tedious process, and thus container orchestration platforms, such as Kubernetes or Docker Swarm, have been developed to help automate deployment and management of containers.[38,39] However, setup and configuration of these orchestration platforms can be complex, and misconfigurations are not uncommon.[40]

To address the above concerns, we posit that an approach for processing large-scale neuroimaging datasets should have the following design criteria: 1.) a data archival method with sufficient storage and security to handle datasets with varying levels of security protocol, 2.) a computational arrangement that is able to scalably process the large amounts of data without extensive financial overhead, 3.) efficiency in processing speed and data transfer of both large and small files, and 4.) deployment of processing pipelines in a reproducible environment with minimal effort and complexity (Table 1-3). Many potential options are not ideal for fulfilling the above design criteria. In short, researchers who curate and manage large neuroimaging datasets have a need for an alternative method of storing and processing data in a reproducible manner that is efficient and cost-effective (Fig. 1).

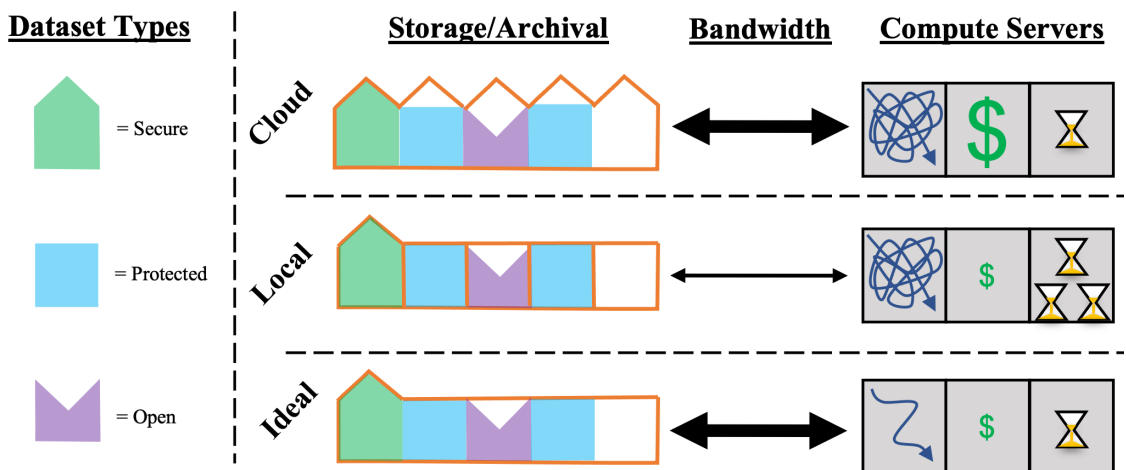

Figure 1. When curating and maintaining large neuroimaging datasets as a team, there are issues with data transfer, data archival, transfer speeds, processing time, cost, and complexity for using different options. While the possibilities of cloud computation permit a large upper bound on all options, the overhead to achieve the scalability is often too large. Further, automation of processing on the cloud can require complexity in setup. However, most local options are plagued with either low bandwidth and slow processing speeds that prevent scalability to large datasets or require a more complex setup with regard to access, permissions, and multiple filesystems. In an ideal scenario ("adaptive"), the compute efficiency and bandwidth are both high, while keeping costs and complexity relatively low.

For groups with access to both an HPC and local storage that share a high-bandwidth network, we propose an alternative strategy for management and processing of large-scale data that adheres to the above design criteria. We maintain a multi-site structural and diffusion-weighted magnetic resonance imaging (MRI) database consisting of 20 large scale datasets, which require various levels of safety and compliance measures. Our data processing consists of 16 separate pipelines that are computationally and time intensive, all of which are contained within Singularity images. The computation is parallelized using the resources of the Advanced Computing Cluster for Research and Education (ACCRE) at Vanderbilt University, which also handles process management and scheduling.

## 2. METHODS

We organize all datasets according to the Brain Imaging Data Structure (BIDS) (v.1.9.0) format.[41] As storage space is limited, we do not organize or process other scan types, such as functional MRI, even if the data are available to download for any given dataset. Our processing system consists of several pipelines that perform various image processing tasks, such as image artifact correction, segmentation, surface recovery, atlas-based registration, and tractography among others. The processing and organization are managed by a team of researchers, with the data available for multiple collaborators to use.

### 2.1 Data Acquisition and Organization

We plan our system to maintain a neuroimaging database consisting of both T1-weighted (T1w) images and diffusion-weighted images (DWI) for research and analysis of diffusion MRI data. Images are received in either the Neuroimaging Informatics Technology Initiative (NIFTI)[42] or Digital Imaging and Communications in Medicine (DICOM)[43] format, where we select DICOM if given a choice. For any DICOMs or NIFTIs that are corrupted or missing information, we ask the providers of the data for complete versions if possible. We then convert DICOMs to NIFTI format using dcm2niix, which also produces a JSON sidecar with metadata information.[44] After conversion, scans are filtered based on protocol, image resolution, image matrix dimensions, and fast visual QA to separate out T1w and DWI images.

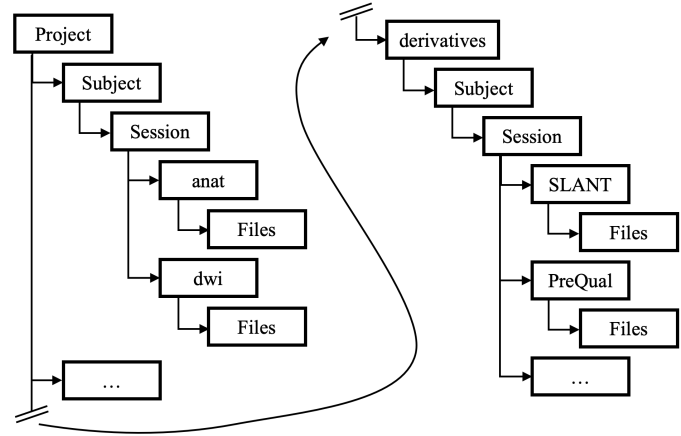

Figure 2. Our file structure uses the BIDS format and validated using the Python version of the BIDS validator. For format consistency with widely used pipelines, we keep our post-processed data in the same output format native to each individual process uses. As we run a multitude of pipelines on our data, keeping the outputs inside separate structured pipeline directories helps with organization and user-friendliness. Additionally, many pipelines require multimodal input, so we have removed the "anat", "dwi", and other modality specific organization directories in the "derivatives" subfolder to avoid confusion.

Files for a dataset are then organized into the BIDS format and validated with the Python version of the BIDS validator tool.[41,45] We aim to maintain as much of the original session naming and identification structure as possible when constructing BIDS-compliant naming schemes in order to allow files to be easily mapped back to original identifiers. All datasets are kept as separate directories in a parent folder. As a small added measure of security, the BIDS-organized files inside dataset directories are all symbolic links to the raw and processed data files that exist outside the BIDS-organized folders. Downstream data obtained through processing pipelines is kept in a separate "derivatives" directory at the top level of the BIDS dataset directory (Fig. 2). As we run several widely used processing pipelines on our data, such as Freesurfer[46], SLANT[47,48], UNesT[49], and PreQual[50], we maintain the original output structure under the "derivatives" folder for each process. Currently, our database consists of 20 separate datasets acquired from various national-scale studies and research initiatives (Table 4). We are continuing to receive, curate, and process more data as they become available. For studies that continue to scan participants, such as the Alzheimer's Disease Neuroimaging Initiative (ADNI) or the National Alzheimer's Coordinating Center (NACC) that continue to scan participants, we pull new scans on a 6-to-12-month basis.

**2.2 Technological Resources**

For computation, we use the Advanced Computing Center for Research and Education (ACCRE) cluster at Vanderbilt University, which consists of 750 compute nodes, 20,100 CPU cores, and almost 200 Terabytes (TB) of RAM.[51] For researchers, the on-demand cost is $84/core/year,[52] with a discount on pricing for purchasing compute time up-front (called "fairshare") rather than on-demand. Backed-up ACCRE filesystem storage is priced at $180/TB/year, which can result in $72,000 per year for 400TB of storage. Rather than pay this high cost for a multi-year project, we house most of our data on a 407 TB server with RAID-Z2 configuration; however, some data that require additional General Data Protection Regulation (GDPR) compliance are stored on a separate GDPR compliant server with 266 TB of storage and RAID-Z2 configuration. The network within the ACCRE cluster, including the storage servers, can reach data transfer speeds of up to 100 Gigabits (Gb) per second. The resources of the ACCRE cluster are shared by many schools and research groups, and management of resource partitioning is done through the Simple Linux Utility for Resource Management (SLURM).[53] Data are backed up nightly to an Amazon Glacier Deep Archive with dynamic storage space that costs $0.0036 Gigabytes (GB) per month for storage space.[54] These choices for computation and storage allow us to further improve cost-efficiency while adhering to design criteria of proper data archival, efficient data transfer, and scalable data processing.

**2.3 Data Processing Management**

For running processing pipelines on our datasets, we employ a semi-automated approach designed to ensure reproducibility, efficiency, and adherence to safe practices (Fig. 3). All pipelines are containerized as Singularity image files, as Singularity permits any user to run processing regardless of admin permissions.[34] The Singularity image files are stored in a separate archive that is accessible to any computation node on the ACCRE cluster. Upon a user specifying a dataset and pre-/post-processing analysis to run, the data archive is automatically queried for data that is available to run but has not yet been run through the analysis. Individual process scripts are then generated for each data instance, and a SLURM job array script is also generated according to specifications the user provides. An accompanying CSV file is output that indicates which scanning sessions in the dataset did not meet the criterion for a processing pipeline, for example, no available T1w image in the scanning session, and what the cause was.

The user then can submit jobs for processing to the ACCRE cluster, which handles job scheduling. As storage on computation nodes is limited and expensive, the input files are copied from the database to the compute nodes inside the job scripts. During the job, the script executes the appropriate containerized Singularity image similar to the "spiders" in Huo et al.[26] On completion of the pipeline, outputs are copied back to the appropriate storage server in the BIDS format. All file transfers that occur are also assessed for data integrity with checksums, with any non-match resulting in the termination of the job script with an error notification. A configuration file is also provided with the outputs that specifies when the process was run, who the user was that ran the process, and the paths to input files used in the analysis for file provenance. For burstable job submission when ACCRE resources are unavailable due to capacity limits or maintenance or general debugging of pipelines, the query and script generation is compatible with any local server as well, with the only difference being a Python file as output that parallelizes processing instead of a SLURM job array.

While our data processing method is automated with respect to archive querying and script generation, users must still decide when to manually run the single line script generation code and submit the processing jobs to ACCRE. Thus, we implement a simple query for both resource usage and storage to inform our team of the current usage status for the cluster and local resources. This automated resource evaluation helps inform our decision-making process in order to maintain the design criterion of efficient data processing.

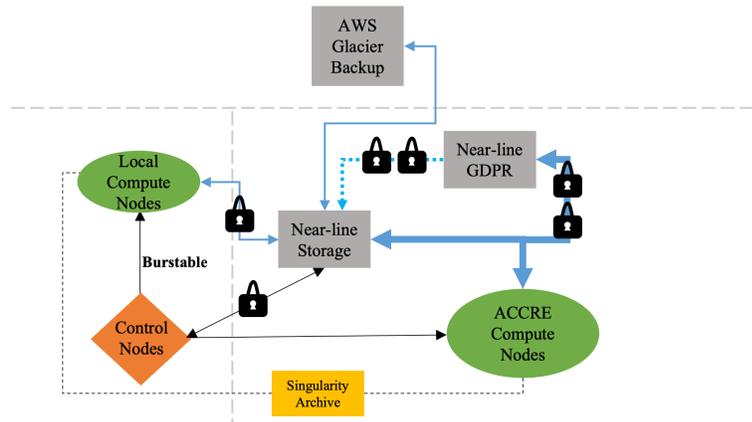

Figure 3. We propose an adaptive solution for scalable managing of large neuroimaging datasets. The storage is split between two separate servers: one for datasets that require additional security compliance measures (such as GDPR), and another for all other datasets. The data on the high-security server is symbolically linked to the general-purpose storage only for users with authorized access. Both the storage servers and all HPC compute nodes share a high bandwidth network are located at the ACCRE cluster at Vanderbilt University, which uses a SLURM scheduler for managing computation jobs. For running processing pipelines on a dataset, we design an automated data query to create processing scripts for all data that can be run through a pipeline and have not already been run successfully. A SLURM job array script is also generated that can easily be submitted for processing. All scripts run containerized code located in a Singularity archive. For instances when quick processing bursts are required or ACCRE is unavailable, our data query and script generator is compatible with local servers as well. To further reduce overhead while maintaining data integrity, data are backed up to an Amazon Glacier storage instance, which is comparatively cheaper than backup under ACCRE. All querying and job submission is done from local control nodes. Blue lines represent data transfer, single arrow black lines represent job submission, double arrow black lines represent query, and dotted lines represent symbolic links. Line thickness represents transfer speed.

## 2.4 Compute Environment Comparison

To demonstrate our adherence to the design criteria of low-cost scalable processing and efficient data transfer, we plan a simple experiment using a small dataset of six T1w scans from the MASIVar dataset.[55] Inside a Singularity container, we run all scans through Freesurfer (v.7.2.0), a well-used pipeline for brain segmentation and cortical reconstruction, on local, HPC (ACCRE), and cloud-based (Amazon Web Services - AWS) computational environments, recording the time it takes for processing to complete.[46] We also measure the time for data transfer to occur (bandwidth) between the storage server and the compute node through copying a 1 GB file 100 times. The latency of each network is also measured with 64-byte packets that are transferred 100 times as well. Finally, we compare the final overhead cost to use the computational resources for each method.

## 3. RESULTS

For running Freesurfer on six T1w scans, the computation time is similar across all three environments, taking 375.5 minutes per image on average for our method of processing compared to 355.2 minutes when using AWS cloud computation resources and 386.0 minutes when using local computation resources (Table 1). The average data transfer speeds from the storage server to the compute server are 0.60, 0.33, and 0.81 Gb/s for HPC, cloud, and local computation environments respectively, while the average latencies are 0.16, 19.56, and 1.64 milliseconds respectively. We also compare the cost per hour of computation time for one 16 GB instance between AWS and ACCRE, with ACCRE being about 20 times less expensive per hour than the cloud service. Overall, the total additional direct cost of computation is calculated as $6.59 for AWS compared to $0.36 for using the ACCRE HPC, a nearly 20-fold decrease in cost.

Table 1. Cost and performance metrics for three different types of computation environments. We estimate a research-quality workstation to cost around $4000 and have a five-year life expectancy, and cost is under the assumption that one job runs per workstation.

| Metric | HPC (ACCRE) | Cloud (AWS t2.xlarge) | Local |
|---|---|---|---|
| Average data throughput from storage to compute server (Gb/s $\pm$ stdev) | 0.60 $\pm$ 0.08 | 0.33 $\pm$ 0.01 | 0.81 $\pm$ 0.01 |
| Latency from storage to compute server – 64 bytes transferred (milliseconds $\pm$ stdev) | 0.16 $\pm$ 0.25 | 19.56 $\pm$ 0.17 | 1.64 $\pm$ 0.25 |
| Cost per hr compute time: single instance (dollars) | 0.0096 | 0.1856[56] | 0.0913 |
| Average time to run Freesurfer (mins $\pm$ stdev) | 375.5 $\pm$ 15.5 | 355.2 $\pm$ 9.6 | 386.0 $\pm$ 10.0 |
| Total overhead cost to run Freesurfer jobs (dollars) | 0.36 | 6.59 | 3.53 |

Table 2. Comparison of different pipeline deployment methods.

| Metric | Singularity[34] | Docker[35] | Kubernetes[38] | BIDS-App[57] | NITRC-CE[58] / Other VMs | Local Install |
|---|---|---|---|---|---|---|
| Specific OS Permissions Required | No | Yes | Yes | Yes | No | No |
| Extensive Setup | No | No | Yes | No | No | No |
| Promotes Reproducible Code | Yes | Yes | Yes | Yes | Yes | No |
| Lightweight | Yes | Yes | No | Yes | No | Yes |

Table 3. Comparison of different data archival solutions.

| Metric | XNAT[59] | COINS[60] | LORIS[61] | NITRC-IR[62] | OpenNeuro[63] | LONI IDA[64] | Datalad[65] | CLI |
|---|---|---|---|---|---|---|---|---|
| Requires credentials to use | No | No | No | No | No | Yes | No | No |
| Potential Data Use conflicts for archival | No | Yes | No | Yes | Yes | Yes | No | No |
| Has flexibility in data organizational structure | No | No | No | No | No | No | Yes | Yes |

## 4. DISCUSSION

Using Singularity containers as our pipeline deployment method allows any team member to run pipelines in a reproducible manner on the ACCRE cluster without administrative OS permissions, while keeping the complexity of batch processing low on our end. Our near-line dual storage server system allows us to perform fast data transfer with low latency while still adhering to the required security measures for all datasets. Finally, we have shown our method of data processing to be more cost effective compared to cloud-based methods. In these ways, we have fulfilled the design criteria for maintaining and processing large-scale neuroimaging data in a reproducible manner.

Based on our experimental results, the data transfer speed and latency from local storage to cloud computation nodes can greatly increase the amount of time spent running processing pipelines, especially given data transfer of larger files (tractography files and high-resolution DWI are several GBs in size) and provenance checks for several small output files that occur throughout processing (Table 1). Although the ACCRE cluster has a Gigabit ethernet, the transfer speeds lower than one Gb/s are likely due to the added time to read from the storage server and write to the compute server, which are hard disk drives rather than the solid-state drives for both the local and AWS instances. We note that it is possible to configure a cloud setup where data transfer is incredibly fast, having storage and computation occur on the same cloud server. Another benefit to such a setup is that data storage would not need to be separated between a GDPR storage server and a non-GDPR server. However, such a setup would require a much larger overhead for an AWS instance with enough cores and to accommodate the additional storage. For example, an AWS instance with 448 cores, which is about one-third the number of cores we currently pay for, and 12288 GB of memory costs over $100 per hour to use.[66] Storage can also be on a separate cloud server, but the data transfer speed would likely not be as fast as the near-line storage with our method. As for the processing time, all three compute environments are relatively similar. However, local computation cannot parallelize efficiently to larger datasets, and slightly faster cloud computation speeds do not justify the almost 20-fold increase in cost. We also note the limited assumption for Table 1 that a workstation runs only a single job instance. However, the number of workstations that would be required to run all processing in a reasonable time would become unmanageable and have a sizeable upfront cost. Thus, neither computational environment satisfies both design criteria of scalability and low overhead cost, while our method does. We do note that for computationally intensive pipelines, the data transfer time is minimal compared to the total processing time.

Moreover, processing tens of thousands of MRI scans (in the hundreds of thousands if you include both modalities) through 16 different processing pipelines can become a huge financial sink, especially when the pipelines can take days to run and require dozens of GBs of RAM. Researchers must also keep in mind that the expected cost would be a baseline, and actual costs would likely be much greater due to processing errors, debugging, and resubmitting failed jobs. Additionally, we only consider AWS for cloud computation, but there are other alternatives, such as Google Cloud or Microsoft Azure, that offer different pricing for similar resources.[67,68] While we have the ACCRE cluster available to use for computation, we also note that other HPC clusters may be more or less expensive. Thus, researchers planning to undergo large scale data processing should consider whether the options available to them would be similarly cost-effective as the option available to us.

One thing to note is that our data archival solution is through the command line interface (CLI). Using the CLI allows for much more control and configuration: for example, using XNAT as a data archival solution would not permit us to store data on two separate servers, nor would it allow us to organize our data into the BIDS format. Also, we acknowledge that database-centric archival solutions, such as OpenNeuro[63] or NITRC-IR[62], are fantastic for sharing data with the greater neuroimaging community. However, such options do not adhere to the design criteria for managing datasets at our scale due to the data transfer speeds, not to mention the required additional security for datasets such as UKBB. We recognize that despite the advantages of using a CLI method of data archival, it may not be ideal for researchers who lack familiarity with the command line. Other archival solutions that can complement the CLI, such as Datalad, can be very useful for tracking changes to the organizational structure of data.[65] While Datalad does not track changes to original files if symbolic links are present (as we do) in the data archive, we note that it can still be used to monitor the higher-level structure rather than individual file contents. We do not incorporate Datalad in our methodology due to the current dynamic state of our database with thousands of updates happening daily, as repetitively scraping the entire file structure to push the changes to a structure that do not remain for long would not be very beneficial. However, we will reconsider incorporating Datalad if our database becomes more stable in the future.

We suggest researchers adopt a method similar to ours if a low-cost solution is desired for large-scale data processing with access to an HPC that has a scheduling protocol, as the setup requires much less manual overhead and provides more usability across a team setting. Using Singularity as the containerization platform allows researchers to take advantage of pre-configured cluster management and scheduling systems that are not specifically tailored for managing containers due to the flexibility in required OS permissions. We choose to use Singularity containers over BIDS-Apps[57] for this reason, despite our data being organized into the BIDS format. For researchers who instead prefer BIDS-Apps and Docker as the pipeline deployment method of choice, we plan to release our processing pipelines as BIDS-Apps. We note that Docker images can be converted to Singularity images using *docker2singularity*.[69] With the ACCRE cluster, the fault-tolerance of computation nodes and scheduling is all handled by ACCRE, making setup low-effort and low-complexity on our end. Further, as we expect to access the backed-up data rarely, if at all, the Glacier Deep Archive permits a low-cost data safety measure for large amounts of data. In addition to the benefit of the low cost-performance ratio of our method, the fast near-line storage is helpful for debugging purposes when running pipelines. While we maintain a neuroimaging database of structural and DWI MRI, our method for large scale data processing can also be adapted to other imaging modalities as well.

Table 4. Size of the datasets currently maintained for our neuroimaging database. Note that datasets are at different levels of processing, so smaller datasets that have been run through all processing pipelines may currently take up more storage space than larger datasets at early processing stages.

| Dataset | Participants | Scanning Sessions | Current Size of Total Data (TB) | Raw MRI Image Files | Total File Count |
|---|---|---|---|---|---|
| ABVIB[70] | 188 | 227 | 0.2 | 284 | 69499 |
| ADNI[16] | 2618 | 11190 | 47 | 25524 | 14550555 |
| BIOCARD[71] | 212 | 504 | 8.4 | 3003 | 1180884 |
| BLSA[72] | 1151 | 3962 | 65 | 19043 | 9356630 |
| CAMCAN[73] | 641 | 641 | 0.4 | 1282 | 36537 |
| HABS-HD[74,#] | 4259 | 6496 | 1.1 | 18675 | 469071 |
| HCP-Aging[75] | 725 | 725 | 15 | 1454 | 1727081 |
| HCP-Baby[76] | 213 | 418 | 2.1 | 1938 | 362416 |
| HCP-Development[77] | 635 | 635 | 2.2 | 1271 | 625552 |
| HCP-Young Adult[78] | 1206 | 1206 | 4.5 | 2253 | 1644656 |
| ICBM[79,80] | 193 | 193 | 2.4 | 1168 | 828946 |
| MAP[81] | 589 | 1579 | 12 | 3158 | 2157929 |
| MARS[82] | 184 | 347 | 2.7 | 694 | 474225 |
| NACC[83] | 5739 | 7831 | 16 | 13312 | 3826519 |
| OASIS3[14] | 992 | 1687 | 8.1 | 8164 | 1375463 |
| OASIS4[84] | 661 | 674 | 4.1 | 3942 | 1202282 |
| ROS[85] | 77 | 127 | 1.0 | 254 | 173564 |
| UKBB[86] | 10439 | 10439 | 79 | 29525 | 18734690 |
| VMAP[17] | 769 | 1805 | 9.6 | 4708 | 2046778 |
| WRAP[87] | 612 | 1625 | 7.1 | 3769 | 1831795 |
| **TOTAL** | **32103** | **52311** | **287.9** | **143421** | **62675072** |

## ACKNOWLEDGEMENTS

**This work has not been submitted for publication or presentation elsewhere.**

This work was supported in part by the National Institute of Health through NIH awards K01-EB032898 (Schilling) and K01-AG073584 (Archer), grant number 1R01EB017230-01A1 (Landman), and ViSE/VICTR VR3029, UL1-TR000445, and UL1-TR002243. This work was supported by the ADSP Phenotype Harmonization Consortium (ADSP-PHC) that is funded by NIA (U24 AG074855, U01 AG068057 and R01 AG059716). This work was conducted in part using the resources of the Advanced Computing Center for Research and Education (ACCRE) at Vanderbilt University, Nashville, TN. We appreciate the National Institute of HealthS10 Shared Instrumentation grant 1S10OD020154-01, and grant 1S10OD023680-01 (Vanderbilt's High-Performance Computer Cluster for Biomedical Research).

---

[#] *Previously the HABLE study*

We would also like to acknowledge Dr. Cailey Kerley for her input and help in editing the manuscript.